\documentclass[sigconf]{acmart}
\AtBeginDocument{%
  }

\copyrightyear{2026}
\acmYear{2026}
\setcopyright{cc}
\setcctype{by}
\acmConference[IDE '26]{3rd International Workshop on Integrated Development Environments }{April 12--18, 2026}{Rio de Janeiro, Brazil}
\acmBooktitle{3rd International Workshop on Integrated Development Environments (IDE '26), April 12--18, 2026, Rio de Janeiro, Brazil}
\acmPrice{}
\acmDOI{10.1145/3786151.3788602}
\acmISBN{979-8-4007-2384-1/2026/04}

\usepackage{booktabs,enumitem}

\makeatletter
\let\old@ps@headings\ps@headings
\let\old@ps@IEEEtitlepagestyle\ps@IEEEtitlepagestyle
\def\confheader#1{%
  \def\ps@headings{%
    \old@ps@headings%
    \def\@oddhead{\strut\hfill#1\hfill\strut}%
    \def\@evenhead{\strut\hfill#1\hfill\strut}%
  }%
  \def\ps@IEEEtitlepagestyle{%
    \old@ps@IEEEtitlepagestyle%
    \def\@oddhead{\strut\hfill#1\hfill\strut}%
    \def\@evenhead{\strut\hfill#1\hfill\strut}%
  }%
  \ps@headings%
}
\makeatother

\begin{document}

\title{Proto-ML: An IDE for ML Solution Prototyping}

\author{Selin Coban}
\orcid{0009-0006-1764-8091}
\affiliation{%
  \institution{Research Group Software Construction \\ RWTH Aachen University}
  \city{Aachen}
  \country{Germany}
}
\email{coban@swc.rwth-aachen.de}

\author{Miguel Perez}
\orcid{0009-0004-2267-0387}
\affiliation{%
  \institution{RWTH Aachen University}
  \city{Aachen}
  \country{Germany}
}
\email{miguel.perez@rwth-aachen.de}

\author{Horst Lichter}
\orcid{0000-0002-3440-1238}
\affiliation{%
  \institution{Research Group Software Construction \\ RWTH Aachen University}
  \city{Aachen}
  \country{Germany}
}
\email{lichter@swc.rwth-aachen.de}

\renewcommand{\shortauthors}{Coban et al.}

\begin{abstract}
Prototyping plays a critical role in the development of machine learning (ML) solutions, yet existing tools often provide limited support for effective collaboration and knowledge reuse among stakeholders. This paper introduces \textsc{Proto-ML}, an IDE designed to strengthen ML prototyping workflows. By addressing key deficiencies such as insufficient stakeholder involvement, limited cross-project knowledge reuse, and fragmented tool support, \textsc{Proto-ML} offers a unified framework that enables structured documentation of prototyping activities and promotes knowledge sharing across projects.

The \textsc{Proto-ML} IDE consists of three extension bundles: prototype implementation, analysis, and knowledge management. These extensions support tasks ranging from evaluating prototype quality against defined criteria to incorporating stakeholder perspectives throughout the development process. Preliminary user feedback suggests that \textsc{Proto-ML} can increase prototyping efficiency and foster more transparent and reusable ML solution development.
\end{abstract}

\ccsdesc[500]{Software and its engineering~Integrated and visual development environments}

\keywords{Integrated Development Environment, Machine Learning, Prototyping}


\maketitle

\section{Introduction}

Prototyping is a long-standing practice in software engineering used to explore design alternatives, reduce uncertainty, and obtain early feedback before full-scale system development. In traditional software prototyping, developers typically rely on established engineering practices, with prototypes focusing on system architecture, user interfaces, or selected functionalities.

In contrast, prototyping of machine learning (ML) solutions is inherently exploratory and highly data-centered. The outcome of an ML prototype depends not only on design choices but also on the availability and quality of data, the selection and configuration of algorithms, the evaluation of model performance, and ultimately the acceptance of the ML prototype among stakeholders. 

Current tool support for ML solution prototyping largely builds on tool environments designed for general-purpose software development. Widely used tools such as Jupyter Notebook, version control systems, and documentation platforms are frequently combined in an ad hoc manner. While effective for quick experimentation, this fragmented tool landscape provides limited support for documentation of abandoned paths, traceability of design decisions or reusing knowledge. As a result, knowledge generated during ML prototyping is often lost, and collaboration across disciplinary boundaries is hindered \cite{NAZIR2024111860, shivashankar2025}.

These observations motivate the need for dedicated tool support tailored to the specific characteristics of ML solution prototyping. In this paper, we present the design and implementation of an IDE for ML solution prototyping (referred to as \textsc{Proto-ML}). This work builds upon and extends our initial results for tool-supported ML solution prototyping, as presented in \cite{aydin2024tool}.

\section{ML Solution Prototyping}

ML solution prototyping is a form of experimental prototyping to quickly evaluate alternative ML models for a given ML problem. The developed \emph{ML prototype} serves as both a demonstrator and an executable artifact, providing a shared, interactive basis for communication between stakeholders and go/no-go decisions.

In this section, we describe essential deficiencies that motivate our work and introduce the ML solution prototyping process.

\subsection{Deficiencies}
\label{sec:challenge}
A literature review conducted by Truss et al. \cite{truss2025human} pointed out that one of the most significant challenges in ML prototyping is the knowledge disparity among involved stakeholders. Further studies highlight that non-technical stakeholders often lack access to key information necessary for a general understanding of ML prototypes \cite{SaeedSKH23,zhangcollab, brennan20,MOE2012853}. They may not know how the ML prototype works, what data and ML models are used, or how trustworthy the results are. In addition, stakeholders often lack insight into why certain design decisions were made. 
We discuss these deficiencies in the area of knowledge reuse and the provision of information to non-technical stakeholders in detail below. 

\begin{figure*}
    \centering
    \includegraphics[width=\linewidth]{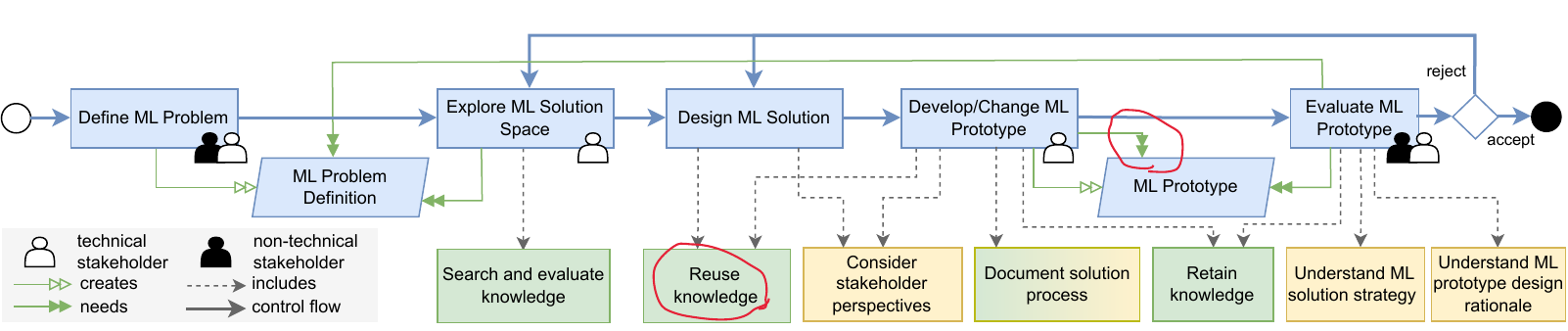}
    \caption{The ML solution prototyping process (adapted from \cite{aydin2024tool})}
    \Description{<long description>}
    \label{fig:extended}
\end{figure*}

\subsubsection*{\textbf{Knowledge is not reused systematically:}}
The knowledge created during the prototyping of an ML solution encompasses all artifacts and insights generated during that process. This covers tangible outputs, such as software components, ML problem-definition documents, and evaluation reports. It also encompasses intangible assets like experiences, lessons learned about data characteristics, and assessments of different solution strategies. These experimental insights are particularly valuable because they can inform future prototyping activities. For example, data scientists may be able to exclude specific approaches from the outset. 

We conducted an interview study \cite{reusestudy} to investigate how reuse is practiced in ML solution prototyping, uncover existing reuse practices, and analyze how tools support these activities, as tools play a key role in accelerating the retrieval and application of reusable artifacts. We observed that developers systematically \textit{search for} and \textit{evaluate knowledge} prior to \textit{reuse the knowledge} and \textit{retain the knowledge}. Notably, the scope of reuse extends beyond code artifacts to encompass ML algorithms, solution design patterns, and step-by-step procedural instructions.


For effective reuse of knowledge developers must \textit{document the prototyping process} and systematically \textit{retain the knowledge} gained to make it available for future prototyping endeavors. Unfortunately, current tool environments do not adequately support the entire knowledge reuse management process. Instead, developers must spend considerable time searching for knowledge, evaluating it, and storing it appropriately.

\subsubsection*{\textbf{Stakeholder-specific information is not provided:}}


ML prototypes are evaluated using a range of quality criteria, which may be quantitative (e.g., accuracy and precision) or qualitative (e.g., fairness and legal compliance). The active involvement of all stakeholders is essential because each brings unique perspectives and distinct concerns regarding the ML problem \cite{brea2023innovative}. Domain experts, in particular, can provide valuable insights into whether the ML prototype will be accepted within its intended application area. It is therefore crucial to \textit{consider stakeholder perspectives} throughout the entire ML solution prototyping process. Stakeholders must receive sufficient and comprehensible information so they can \textit{understand the ML solution strategy} and \textit{understand the design rationale} behind specific design decisions.

\subsection{The ML Solution Prototyping Process}

Figure \ref{fig:extended} depicts the basic ML solution prototyping process in blue. A central artifact is the \textit{ML problem definition}, which specifies the scope, business objectives, and success criteria. Non-technical stakeholders (e.g., domain and business experts) primarily influence the design of this artifact. Guided by the ML problem definition, technical stakeholders (e.g., developers and data scientists) conduct the subsequent activities: \emph{exploring the ML solution space}, \emph{designing the ML solution}, and \emph{developing and changing the ML prototype}. Domain and business experts participate in \emph{evaluating the ML prototype}. Acceptance decisions depend on contextual understanding, anticipated business impact, and ethical and regulatory compliance.

To address the deficiencies in ML solution prototyping described above, we have extended the basic prototyping process to include missing activities that are needed to overcome those deficiencies. On the one hand, these integrate the systematic reuse of knowledge into the prototyping process (shown in green). On the other hand, the extended process includes activities (shown in yellow) that are necessary to better involve non-technical stakeholders, especially in the evaluation of the ML prototype.

%

\section{Tool requirements }
\label{sec:tool-req}
For the proposed ML solution prototyping process to be implemented efficiently, it must be supported by suitable tools. 

Based on the tasks performed in the activities, we formulate requirements for the tool environment aimed at increasing efficiency in the reuse process and addressing the stakeholder information needs. A high degree of automation is required to foster broader acceptance of the solution. We have therefore established the following tool requirements for the newly introduced process activities:

\begin{itemize}
    \item \textbf{Search and evaluate knowledge \& Retain knowledge}\\
    R1: \textit{Provide means to manage knowledge sources, capture their evaluation results, and trace them to the ML prototype.}
    \item \textbf{Reuse knowledge}\\
    R2: \textit{Support fast retrieval of relevant knowledge.}
    \item \textbf{Consider stakeholder perspectives}\\
    R3: \textit{Automatically evaluate ML prototype against defined stakeholder interests.}
    \item \textbf{Develop/Change ML prototype}\\
    R4: \textit{Automatically assess ML prototype against defined quality criteria.}
    \item \textbf{Understand ML solution strategy}\\
    R5: \textit{Provide automated generation of high-level prototype solution overviews or summaries.}\\
    R6: \textit{Provide automated generation of ML prototype documentation tailored to stakeholders.}
    \item \textbf{Understand ML prototype design rationale}\\
    R7: \textit{Provide means to explore the prototyping process.}
    \item \textbf{Document solution process}\\
    R8: \textit{Automatically record the prototyping process and design decisions.}
\end{itemize}


R1–R2 address improving cross-project knowledge reuse. They require mechanisms to manage knowledge sources, support developers in evaluating them, and enable traceability between reused knowledge and the components in the ML prototype to which it corresponds. In addition, since developers currently spend considerable time searching for relevant knowledge, dedicated support for faster retrieval is needed.

R3 and R4 aim to support developers during ML prototype development by checking it against quality attributes and stakeholder interests. This is expected to result in a higher-quality prototype.

To enhance stakeholder understanding of the ML prototype and its prototyping process, R5–R7 focus on the tailored presentation of the prototype’s solution strategy and its prototyping process. 

Finally, R7 is a prerequisite for R8. Without recording the prototyping process, it is not possible to provide information about the process.

Note that this list of requirements is not exhaustive and may be expanded in the future.

\section{Components of the Proto-ML IDE}
All existing tools for prototyping ML solutions and those to be developed based on these requirements should be made available in a consistent prototyping IDE, which we have named \textsc{Proto-ML}.

\begin{figure}
    \centering
    \includegraphics[width=\linewidth]{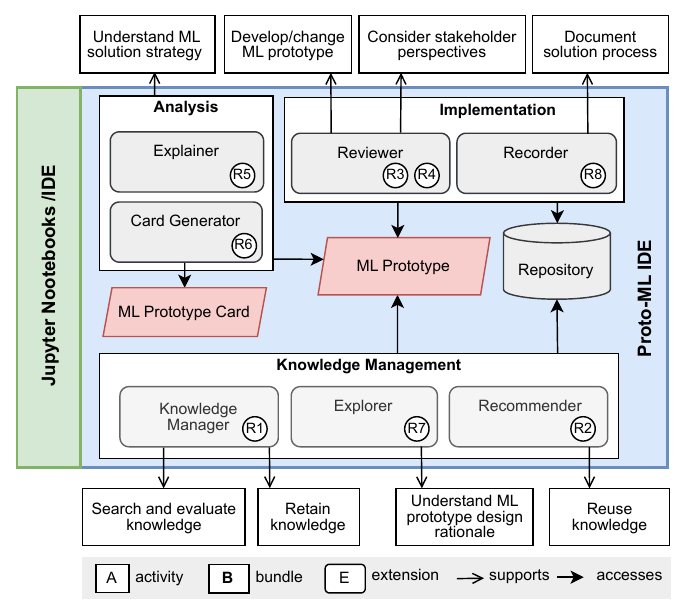}
    \caption{Components of the \textsc{Proto-ML} IDE}
    \Description{<long description>}
    \label{fig:toolbox}
\end{figure}

Jupyter Notebooks, often combined with IDEs such as \textsc{PyCharm} or \textsc{Visual Studio Code}, are the de facto standard for developing ML prototypes \cite{venkatesh2024staticanalysisdrivenenhancements, wang2020better}. For this reason, we have designed the \textsc{Proto-ML} IDE so that Jupyter Notebooks are its central component, with additional bundles of interoperable prototyping-specific tooling extensions. Based on the requirements outlined in Section \ref{sec:tool-req}, we have developed three task-oriented extension packages with a total of seven tools to support ML prototyping. Figure \ref{fig:toolbox} provides an overview of the bundles and extensions of the \textsc{Proto-ML} IDE, along with references to the supported activities and the tool requirements presented. In addition to these bundles, the \textsc{Proto-ML} IDE contains a \textit{repository} in which all data relating to current and previous ML prototypes is stored. 

We implemented proof-of-concept prototypes\footnote{url: \url{https://selincoban95.github.io/proto-ml.github.io/}} using the \textsc{JupyterLab} platform for almost all extensions and evaluated them initially. In the following, we present a first glance at these bundles. Note that some extensions are still under development.

\subsection{Bundle: Analysis}
To better understand the solution strategy implemented in an ML prototype, we suggest two extensions.

\subsubsection*{\textbf{Explainer:}}
\begin{figure}
    \centering
    \includegraphics[width=0.9\linewidth]{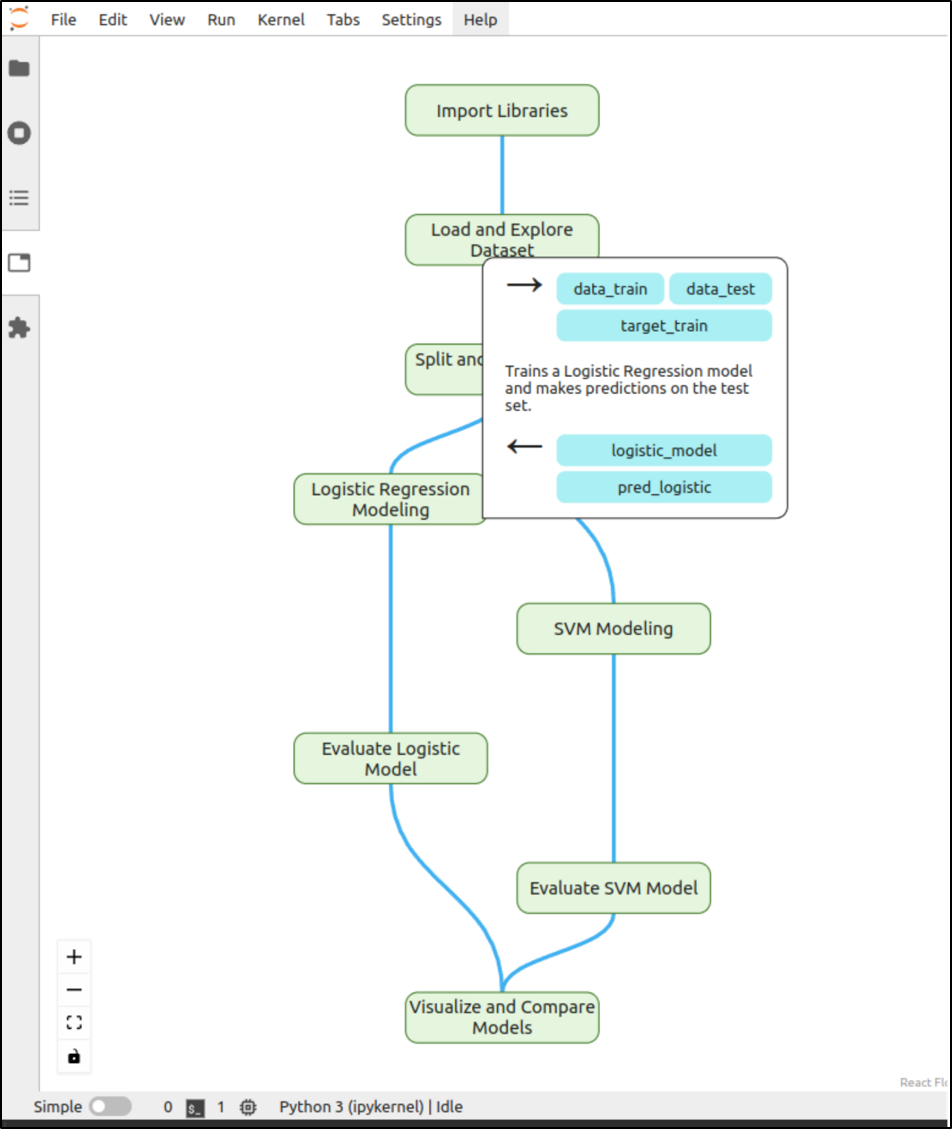}
    \caption{Screenshot of the Explainer extension}
    \Description{<long description>}
    \label{fig:cellflow}
\end{figure}

The Explainer provides a high-level overview of the activity-flow, data-flows between activities, and basic explanations. This helps stakeholders understand the general structure of the ML prototype. To obtain a high-level view of a notebook, each cell represents a semantic activity. By applying heuristics over inter-cell data dependencies, activity-flow diagrams can be derived. Activity names and concise textual descriptions are automatically generated by LLMs. Finally, the Explainer provides a visualization that enables exploration of both the activities and the corresponding data flow. A screenshot of an activity-flow diagram for a sample notebook is provided in Figure \ref{fig:cellflow}.

We collected initial feedback from five ML practitioners with varying technical backgrounds. All participants reported that the Explainer helped them understand given notebooks more quickly and navigate their own notebooks during development. One participant also indicated their intention to use the Explainer to ensure the notebook reflects their intended implementation.

\subsubsection*{\textbf{Card Generator:}}
Further details about an ML prototype are stored in an \textit{ML prototype card} inspired by model cards \cite{mitchell2019model}, an established approach for documenting essential information about ML models. We aim to adapt this concept for ML prototypes and tailor it the information needs of technical and non-technical stakeholders. The Card Generator creates these ML prototype cards semi-automatically. This extension is currently under development.


\subsection{Bundle: Implementation}
In this bundle, we introduce two extensions designed to support developers in creating high-quality ML prototypes that incorporate stakeholder perspectives, as well as an additional extension aimed at automatically documenting the prototyping process.

\subsubsection*{\textbf{Reviewer:}}
\begin{figure}
    \centering
    \includegraphics[width=\linewidth]{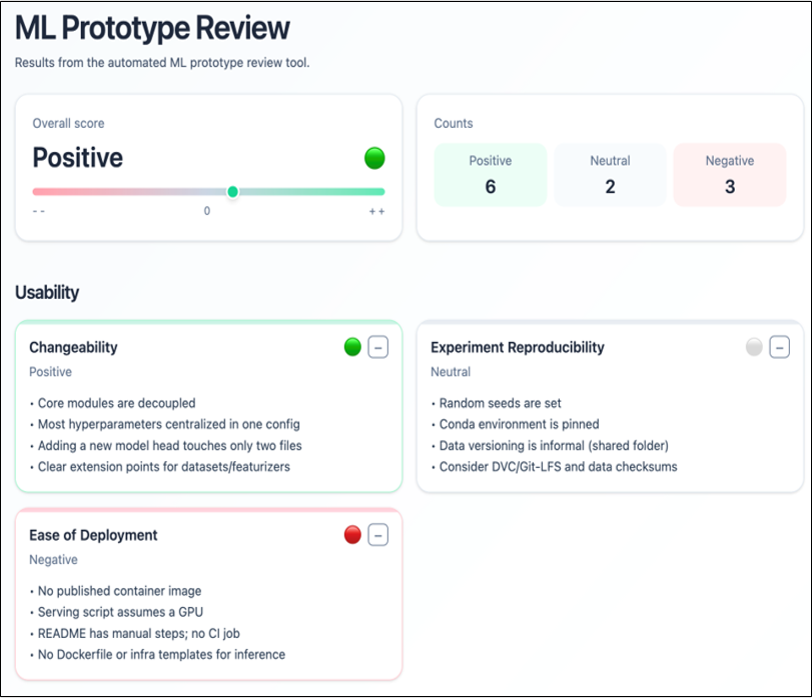}
    \caption{Screenshot of the Reviewer extension dashboard}
    \label{fig:reviewer}
\end{figure}

Similar to static code quality analyzers, we include a Reviewer that evaluates the current state of the ML prototype against ML prototype-specific quality criteria (see Figure \ref{fig:reviewer}). To this end, we developed three artifacts that provide these criteria: (1) a \textit{quality model}, (2) a \textit{review checklist}, and (3) \textit{stakeholder personas}. The quality model was synthesized by combining established quality models for ML systems \cite{qualitymodel, qualityfra} and best-practice guidance for notebooks \cite{qualityjn}. The review checklist contains items to be considered in notebooks (e.g., include a brief description of the ML problem at the top of the notebook), drawing inspiration from the ML prototype card. To incorporate stakeholder perspectives, we designed \textit{stakeholder personas}. For each persona, we identified the specific requirements of stakeholders based on their role and intentions. Together, these artifacts enable an automated, systematic, and stakeholder-oriented evaluation of the ML prototype.

\subsubsection*{\textbf{Recorder:}}
To document the prototyping process, we include a Recorder that automatically tracks all prototyping steps performed in the notebook by storing snapshots. Storing a snapshot of the notebook after every minor modification would quickly lead to excessive amounts of stored data. To address this, we capture snapshots only for semantically meaningful modifications by treating the execution of a single cell as an atomic activity, such as data processing or model training. The stored snapshots enable time-travel across different versions of the ML prototype. While linear version traversal is already supported by existing tools such as \textsc{Git} or \textsc{Verdant} \cite{messyHistory}, we introduced the concept of an \textit{experiment tree} \cite{histree}: When a developer navigates to an earlier version of the notebook and modifies it, a new branch is created in the version history. This branching mechanism allows developers to capture alternative solution paths in a manner consistent with the exploratory nature of rapid experimentation.

\subsection{Bundle: Knowledge Management}
Since existing tools do not yet support knowledge reuse in ML solution prototyping, we include extensions to support searching, evaluating, reusing, and retaining knowledge.

\subsubsection*{\textbf{Explorer:}}
\begin{figure}
    \centering
    \includegraphics[width=0.9\linewidth]{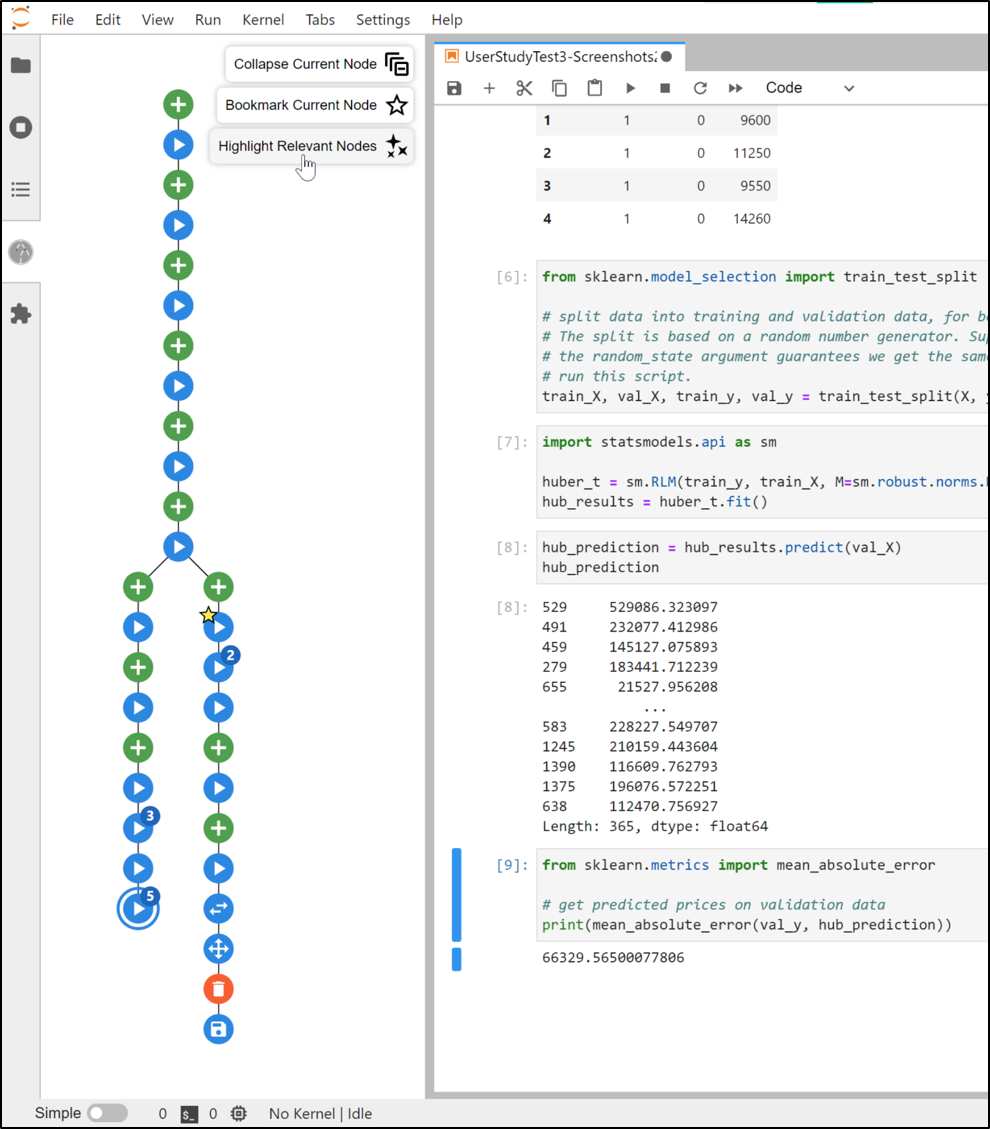}
    \caption{Screenshot of the Explorer extension}
    \Description{<long description>}
    \label{fig:histree}
\end{figure}

The Explorer presents data from the recorded prototyping process to a stakeholder and allows exploration of past prototyping endeavors, along with comments on design decisions. Using the experiment tree generated by the Recorder, the Explorer provides a tree-based visualization of the notebook’s evolution  (see Figure \ref{fig:histree}). Each node in the tree represents a snapshot of the notebook at a specific point in time. When a developer selects a node, the corresponding version is opened. If the developer modifies this version, a new branch is automatically created in the experiment tree. By hovering over a node, additional contextual information becomes available, such as a diff view highlighting changes and any developer comments associated with that snapshot. This design allows developers to rapidly switch between experiment variants and inspect what has already been tried, within a notebook. More details on our Explorer implementation, along with the results of user experiments, are published in \cite{histree}. 

\subsubsection*{\textbf{Recommender:}}
\begin{figure}
    \centering
    \includegraphics[width=\linewidth]{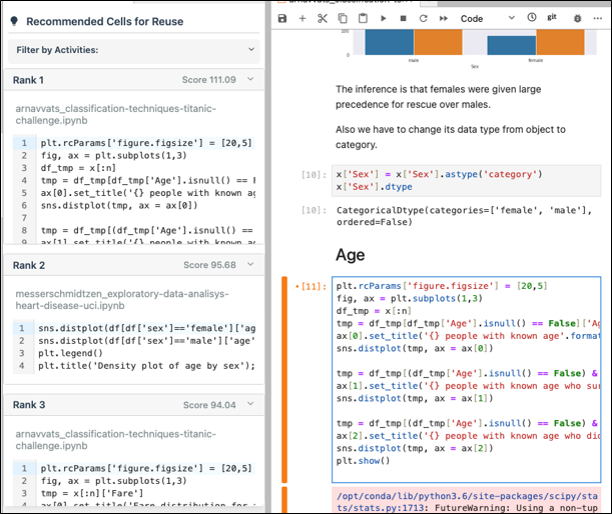}
    \caption{Screenshot of the Recommender extension}
    \Description{<long description>}
    \label{fig:recsys}
\end{figure}

Because notebook code cells are the most frequently reused artifact across notebooks, we developed a recommendation system that operates at two granularity levels: (i) notebook-level, suggesting notebooks similar to the one currently open; and (ii) cell-level, suggesting cells similar to the developer’s active cell  (see Figure \ref{fig:recsys}). The Recommender leverages a vector database populated with representations derived from personal notebooks and a corpus of Kaggle notebooks \cite{kgtorrent} to enable content-based retrieval. Although offline processing (e.g., training and indexing) can be time-consuming, online inference is low-latency, delivering recommendations interactively during editing. 

More details on the Recommender and its performance evaluation can be found in \cite{aydin2024automated}. Applying standard recommender system metrics, the Recommender consistently achieved strong performance. The analysis also identified characteristics that influence recommendation quality, guiding on when and how the Recommender is most effective and informing future design and optimization.

\subsubsection*{\textbf{Knowledge Manager:}}
Using the Knowledge Manager, developers can store and retain various knowledge sources during exploration of the solution space and trace parts of the ML prototype to the corresponding knowledge sources. Furthermore, the Knowledge Manager includes mechanisms to automatically evaluate the suitability of a knowledge source based on criteria identified in an interview study \cite{reusestudy}. For example, it analyzes the source and provides information on whether code is available, whether the author is known and reputable, and whether standard benchmarks were used. As this extension is still in the design phase, further technical and functional details cannot be disclosed at this stage.

\section{Related Work} 
A variety of tools and techniques have been proposed to support isolated aspects of ML solution prototyping, such as code reuse or debugging. A comprehensive review of these techniques and tools around the Jupyter Notebook environment is provided in \cite{siddik2025}. In the following, we present some notable tools in the areas of ML prototype analysis, implementation, and knowledge management, structured according to the extension bundles presented in this paper.

\subsubsection*{Analysis}
Bhat et al. \cite{bhat2023aspirations} introduce \textsc{DocML}, a tool that integrates model cards directly into notebook environments. It assists developers by reminding them of incomplete model card entries and enabling the specification of relevant information through special HTML comments embedded within the corresponding code cells. However, an empirical study by Bracamonte et al. \cite{bracamonte23} found that the information presented in model cards is often perceived as overly technical, potentially excluding non-technical stakeholders. To address this gap, our Card Generator tailors model cards to the prototyping context by aligning their content with the specific information needs of different stakeholders.

The idea of using visual non-linear workflow models to support stakeholders in understanding ML prototypes was also explored by Ramasamy et al. \cite{ramasamy2023visualising}. The authors evaluated the idea with a visual concept and demonstrated its potential effectiveness in improving comprehension. However, the presented idea was never implemented and evaluated in practice.

\subsubsection*{Implementation}
To support structured development in notebooks, best practices have been proposed \cite{pimentel2019large}, and a static notebook analyzer \textsc{Pynblint} has been introduced by Quarnata et al. to automatically check compliance with a subset of these practices aimed at improving collaboration \cite{quaranta2022pynblint}. These best practices are also integrated into the ML prototype's quality model within the Reviewer. 

Recording a linear history of notebook versions is supported by the tools presented in \cite{foraging, nextj}. The Recorder builds on the features of \textsc{Verdant} \cite{foraging} and extends them to support the recording of tree-based histories.

\subsubsection*{Knowledge Management}
Several tools assist developers in retrieving relevant code from past notebooks. One approach is by providing better search support, e.g., through semantic search \cite{li2021nbsearch, li2024unlockinginsightssemanticsearch} or complex graph-based queries \cite{jupysim}, which both require manual effort from the developer. Existing automatic recommendation approaches are either only tailored to specific tasks, such as data analysis \cite{Watson2019PySnippetAE, edaassist}, or consider only existing markdown cells in the notebook \cite{typhon}.

\section{Conclusion and Future Work}

In this paper, we introduce \textsc{Proto-ML}, an IDE for ML solution prototyping that addresses key deficiencies such as limited stakeholder participation and low reuse potential of prototyping knowledge. We present an enhanced process model for ML solution prototyping that explicitly incorporates activities to mitigate these deficiencies. From this process model, we derive concrete tool requirements to support the identified activities effectively.

To realize these requirements, we build on the \textsc{JupyterLab} platform and present three extension bundles that support the analysis, implementation and knowledge management of ML prototypes. Our preliminary evaluation of these extensions has received positive feedback from users, indicating their potential to improve prototyping efficiency.

Several extensions of \textsc{Proto-ML} are still under development and need to be completed. Once the full IDE is available, we plan to conduct empirical studies in real-world ML prototyping projects to evaluate its impact on knowledge reuse and stakeholder engagement.

\bibliographystyle{ACM-Reference-Format}
\bibliography{bibliography}

\end{document}